\documentclass[twocolumn,secnumarabic,amssymb, nobibnotes, showpacs, aps, prl]{revtex4-1}
\usepackage{graphicx}

\usepackage{epsfig}

\usepackage{lineno}
\begin{document}
\newcommand{\sqrtsNN}{\mbox{$\sqrt{\mathrm{s}_{_{\mathrm{NN}}}}$}}
\title{The Proton-$\Omega$ correlation function in Au+Au collisions at \sqrtsNN\,\,=200 GeV}

\author{
J.~Adam$^{11}$,
L.~Adamczyk$^{1}$,
J.~R.~Adams$^{34}$,
J.~K.~Adkins$^{24}$,
G.~Agakishiev$^{22}$,
M.~M.~Aggarwal$^{36}$,
Z.~Ahammed$^{56}$,
N.~N.~Ajitanand$^{48}$,
I.~Alekseev$^{2,30}$,
D.~M.~Anderson$^{50}$
% ...
}

\address{$^{1}$AGH University of Science and Technology, FPACS, Cracow 30-059, Poland}
\address{$^{2}$Alikhanov Institute for Theoretical and Experimental Physics, Moscow 117218, Russia}
% ...
\address{$^{11}$Creighton University, Omaha, Nebraska 68178}
% ...
\address{$^{22}$Joint Institute for Nuclear Research, Dubna 141 980, Russia}
% ...
\address{$^{24}$University of Kentucky, Lexington, Kentucky 40506-0055}
% ...
\address{$^{30}$National Research Nuclear University MEPhI, Moscow 115409, Russia}
% ...
\address{$^{34}$Ohio State University, Columbus, Ohio 43210}
% ...
\address{$^{36}$Panjab University, Chandigarh 160014, India}
% ...
\address{$^{48}$State University of New York, Stony Brook, New York 11794}
% ...
\address{$^{50}$Texas A\&M University, College Station, Texas 77843}
% ...
\address{$^{56}$Variable Energy Cyclotron Centre, Kolkata 700064, India}
% ...


\begin{abstract}
We present the first measurement of the proton-$\Omega$ correlation function in heavy-ion collisions for central (0-40$\%$) and peripheral (40-80$\%$) Au+Au collisions at \sqrtsNN\,\,=200 GeV by the STAR experiment at the Relativistic Heavy-Ion Collider (RHIC). Predictions for the ratio of peripheral collisions to central collisions for the proton-$\Omega$ correlation function are sensitive to the presence of a nucleon-$\Omega$ bound state.  These predictions are based on the proton-$\Omega$ interaction extracted from (2+1)-flavor lattice QCD calculations at the physical point. The measured ratio of proton-$\Omega$ correlation function from peripheral (small system) to central (large system) collisions is less than unity for relative momentum smaller than 40 MeV/\textit{c}. Comparison of our measured correlation ratio with the theoretical calculation slightly favors a proton-$\Omega$ bound system with a binding energy of $\sim$ 27~MeV. 
\end{abstract}
\maketitle

\section{INTRODUCTION}

The study of nucleon-nucleon (NN), hyperon-nucleon (YN) and hyperon-hyperon (YY) interactions are of fundamental importance in understanding relativistic heavy-ion collisions~\cite{RHIC1, RHIC2}, modeling of neutron stars~\cite{neustar1, neustar2, neustar3, neustar4} and examining the existence of various exotic hadrons~\cite{Jaffe, XiXi,DiOmega}. A significant amount of NN scattering data acquired over the years allows us to construct precise NN potential models~\cite{NN_Potential, Lattic1}. The availability of nominal YN scattering data and no scattering data for the multi-strange YY systems makes the task of constructing YN and YY potentials very challenging. With the development of sophisticated computational techniques, it has become possible to carry out first principle calculations based on lattice Quantum Chromodynamics (QCD) to provide constraints on some of the NN, YY and YN interactions~\cite{Lattice1, Lattice2, Lattice3, Lattice4,Nomega_BE2}. Very often the experimental information on the bound states of strange baryons and nucleons (hypernuclei) is used to provide information on YN interactions~\cite{STARHT1,STARHT2,AGal}. However, this method becomes difficult to use because these measurements are contaminated by many-body effects, which makes it very difficult to extract N$\Xi$, N$\Omega$, Y$\Xi$ and Y$\Omega$ interactions.

High-energy heavy-ion collisions produce a sizable number of hyperons in each collision~\cite{strangebaryon}, which provides an excellent opportunity to study the NN, YN and YY interactions. Measurement of two-particle correlations at low relative momentum, also known as femtoscopy, have been used to study the space-time dynamics of the source created in heavy-ion collisions. In addition to this, the  measurement of two-particle correlations at low relative momentum can also be used to measure final state interactions (FSI) between NN, YN and YY. This approach has been used by the STAR experiment at RHIC to extract the FSI for $\Lambda\Lambda$~\cite{LmLm_STAR} and antiproton-antiproton~\cite{pbar_STAR}. 
 
Recent study of (2+1)-flavor lattice QCD simulations for heavy quark masses shows that the nucleon-$\Omega$ interaction (N$\Omega$) is attractive at all distances~\cite{Nomega_BE2}. Using this N$\Omega$ interaction, it is shown that the shape of the two particle correlation function at low relative momentum changes substantially with the strength of the N$\Omega$ attraction~\cite{KM_AO}. However, the presence of the Coulomb interaction  in the  proton-$\Omega$ channel makes it difficult to access the strong interaction directly from the measured two-particle correlation function. Therefore, a new measure, namely the ratio of the correlation functions between the peripheral (small) and central (large) collision systems is proposed in  Ref~\cite{KM_AO}. This ratio provides direct access to strong interaction between proton and $\Omega$, independent of the model used for the emission source. 

The attractive nature of an N$\Omega$ interaction leads to the possible existence of the N$\Omega$ dibaryon with strangeness~$=$~-3, spin~$=$~2, and isospin~$=$~1/2, which was first proposed in~\cite{quark_pot}.  Such an N$\Omega$ dibaryon is the most interesting candidate~\cite{quark_pot, Skyrme, NOmega2, NOmega3, NOmega4} after the H-dibaryon~\cite{Jaffe}. The Pauli exclusion principle does not apply among quarks in the  N$\Omega$ dibaryon  and it is stable against strong decay~\cite{CC1, CC2}.  Several attempts have been made to estimate the binding energy of the  N$\Omega$ state in different QCD motivated models~\cite{Nomega_BE, Nomega_BE2}. The N$\Omega$ dibaryon can be produced in high-energy heavy-ion collisions through the coalescence mechanism~\cite{Yield_NS}. For an S-wave bound state of nucleon and $\Omega$, the strong decays to octet-decuplet systems are prohibited by kinematics and those into octet-octet systems (e.g. $\Lambda\Xi$) are suppressed dynamically due to the D-wave nature~\cite{Nomega_BE2}. This makes direct searches via the invariant mass method very challenging in heavy-ion collisions. The measurement of  the proton-$\Omega$ correlation function for peripheral and central Au+Au collisions at $\sqrt{s_{NN}}=200$ GeV, presented in this Letter, will provide insight into the existence of an N$\Omega$ dibaryon.

\section{DATA ANALYSIS}

STAR is a large acceptance detector at RHIC~\cite{STAR}. The measurements presented in this Letter are from the data taken for Au+Au collisions at \sqrtsNN\,\,=200 GeV in 2011 and 2014. $ 5.30\times 10^8$ minimum bias events from 2011 and $8.76\times 10^8$ minimum bias events from 2014 were analyzed. The tracking and particle identification for the measurements were provided by the Time Projection Chamber (TPC)~\cite{STAR_TPC} and Time-of-Flight (TOF)~\cite{STAR_TOF} detectors. These detectors are located in a 0.5 T magnetic field, which allows determination of the momentum and charge of the particles traversing the TPC. Minimum bias triggered events were selected by requiring coincident signals at forward and backward rapidities in the Vertex Position Detectors (VPD)~\cite{STAR_VPD} and requiring a signal at mid-rapidity in the TOF.  Centrality was determined by the charged particle multiplicity at mid-rapidity ($|\eta|<$0.5) in the TPC. To suppress events from collisions with the beam pipe, the reconstructed primary vertex was required to lie within a 2~cm radial distance from the center of the beam pipe. In addition, the $z$-position of the vertex was required to lie within $\pm 40$ and $\pm 6$~cm of the center of the detector for the data from years 2011 and 2014, respectively.

\subsection{$\Omega$ IDENTIFICATION}
\begin{figure*}
\begin{center}
\epsfxsize = 6.0in
\epsfysize = 3.5in
\epsffile{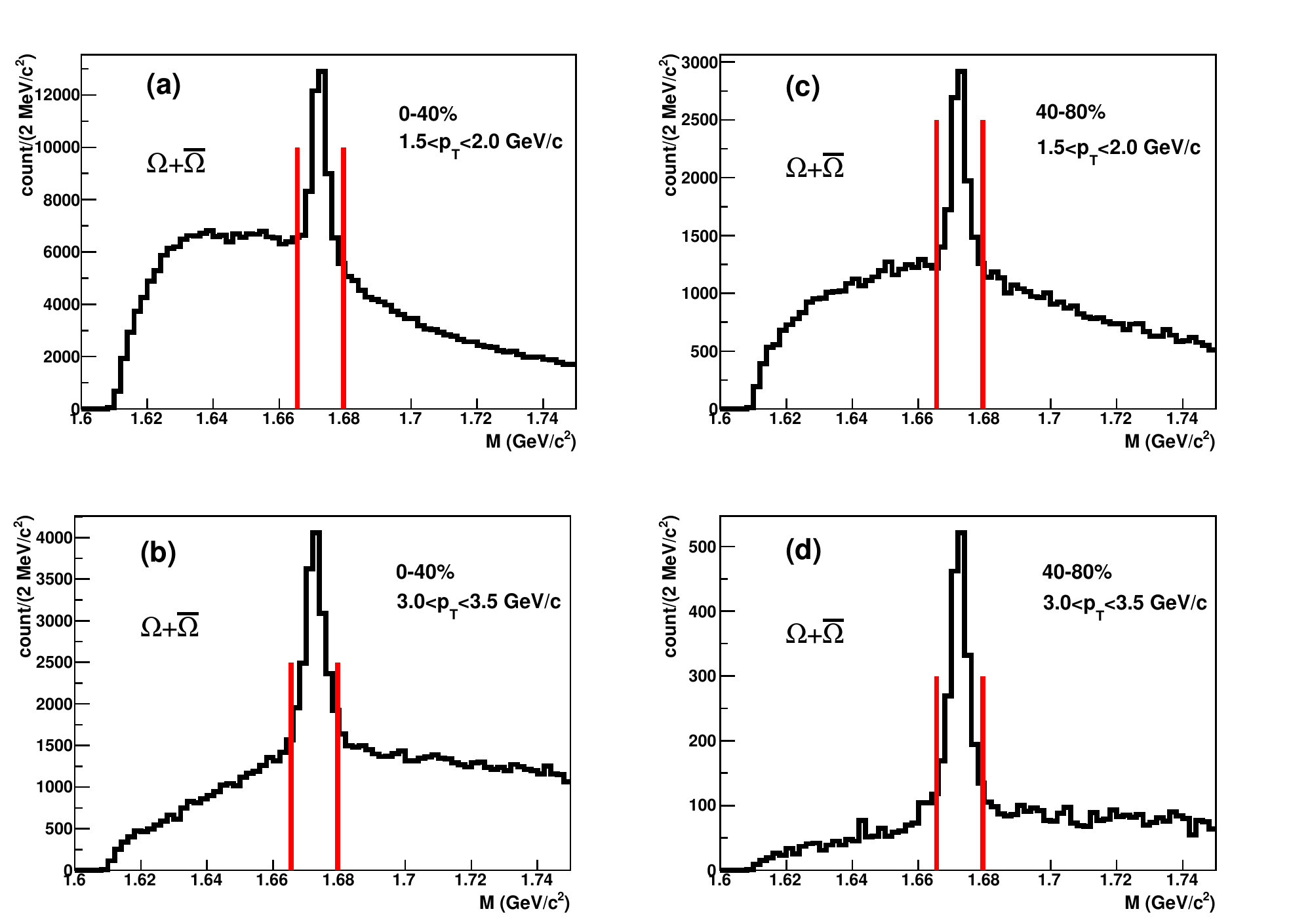} 
\end{center}
\caption{ (color online). Reconstructed invariant mass (M) distributions of combined $\Omega$ and $\bar{\Omega}$ sample for 0-40$\%$ Au+Au collisions at $\sqrt{\normalfont{s_{NN}}}$=200 GeV for the transverse momentum ($p_{T}$) range 1.5$<p_{T}<$2.0 GeV/\textit{c} (a) and 3.0$<p_{T}<$3.5 GeV/\textit{c} (b).  The invariant mass distributions of combined $\Omega$ and $\bar{\Omega}$ sample for 40-80$\%$ Au+Au collisions at $\sqrt{\normalfont{s_{NN}}}$=200 GeV for the transverse momentum ($p_{T}$) range 1.5$<p_{T}<$2.0 GeV/\textit{c} (c) and 3.0$<p_{T}<$3.5 GeV/\textit{c} (d). The solid lines at 1.665 and 1.679 GeV$/c^2$ show the mass region of the reconstructed $\Omega$ and $\bar{\Omega}$ candidates used for the measurement of the proton-$\Omega$ correlation function.}
\label{fig:OmegaReco}
\end{figure*}

The TPC was used for tracking, decay topology and identification of particles for $\Omega$ ($\bar{\Omega}$) reconstruction in the pseudorapidity range $|\eta|< 1$. To reconstruct the $\Omega$ ($\bar{\Omega}$), the decay channel $\Omega(\bar{\Omega}) \rightarrow \Lambda K^{-} (\bar{\Lambda} K^{+})$, with a branching ratio of 67.8\%, with subsequent decay $\Lambda(\bar{\Lambda})  \rightarrow p\pi^{-} (\bar{p}\pi^{+})$ (branching ratio of 63.9\%) was used~\cite{PDG}. The $\Lambda$ ($\bar{\Lambda}$) candidates were formed from pairs of $p$ ($\bar{p}$) and $\pi^{-}$ ($\pi^{+}$) tracks whose trajectories pointed to a common secondary decay vertex, which was well separated from the $\Omega$ ($\bar{\Omega}$) vertex.  These $\Lambda$ ($\bar{\Lambda}$) candidates were then combined with bachelor $K^{-}$ ($K^{+}$) tracks, which points to a common decay vertex well separated from the primary vertex. The decay length (DL) of an $\Omega$ ($\bar{\Omega}$) candidate was required to be larger than 4~cm from the primary vertex. As listed in Table~\ref{tab:cuts}, additional selection criteria on the distance of closest approach (DCA) between the two $\Lambda$ ($\bar{\Lambda}$) daughter tracks, between the $\Lambda$ ($\bar{\Lambda}$) and bachelor track, the $\Lambda$ ($\bar{\Lambda}$) and the primary vertex position were applied to select $\Omega$ ($\bar{\Omega}$). Furthermore, the pointing angle of $\Omega$ ($\bar{\Omega}$) track with respect to the primary vertex ($\vert(r_{\Omega}-r_{PV})\times p_{\Omega}\vert$/$\vert r_{\Omega}-r_{PV}\vert\vert{p_{\Omega}}\vert$, where {\it{r}} is the position of $\Omega$ and primary vertex, respectively and $p_{\Omega}$ is the momentum of $\Omega$) were applied to select $\Omega$ ($\bar{\Omega}$). To reduce the combinatorial background, $\Lambda$ ($\bar{\Lambda}$) candidates were selected in the invariant mass range between 1.112 and 1.120 GeV$/c^2$.  In addition, the candidates due to misidentification of $\pi^{-}$ ($\pi^{+}$) tracks as the bachelor $K^{-}$ ($K^{+}$) tracks were removed by checking a $\Xi$ hypothesis. The invariant mass distributions of combined $\Omega$ and $\bar{\Omega}$ candidates for  0-40$\%$ and 40-80$\%$ Au+Au collisions at \sqrtsNN\,\,=200 GeV  for the  transverse momentum ($p_{T}$) ranges 1.5$<p_{T}<$2.0 GeV/\textit{c} and 3.0$<p_{T}<$3.5 GeV/\textit{c} are shown in Figure~\ref{fig:OmegaReco}(a-d). The signal ($S$) to signal+background ($S+B$) ratio, integrated over $\pm3\sigma$, is 0.2  for the $p_{T}$ range 1.5$<p_{T}<$2.0 GeV/\textit{c} and 0.4  for the $p_{T}$ range 3.0$<p_{T}<$3.5 GeV/\textit{c} in 0-40\% centrality and is 0.3  for the $p_{T}$ range 1.5$<p_{T}<$2.0 GeV/\textit{c} and 0.7  for the $p_{T}$ range 3.0$<p_{T}<$3.5 GeV/\textit{c} in 40-80\% centrality.  All candidates with invariant mass between 1.665 and 1.679 GeV$/c^2$ were used in the analysis.

\begin{table*}
  \centering
  \begin{tabular}{  c | c | c | c  }
    \hline
    Selection criteria &\multicolumn{2}{c|}{0-40\%} &40-80\% \\
\cline{2-3}
    & $p_{T} < 2.5$ GeV/c  & $p_{T} > 2.5$ GeV/c  & All  $p_{T}$ \\
    \hline
    $\Omega$ DCA & $<$ 0.6 cm & $<$ 0.7 cm & $<$ 0.8 cm\\
    $\Lambda$ DCA & $>$ 0.4 cm & $>$ 0.3 cm & $>$ 0.3 cm\\
    DL($\Omega$) & $>$ 4.0 cm & $>$ 4.0 cm & $>$ 4.0 cm\\
    DL($\Lambda$) & $>$ 6.0 cm & $>$ 6.0 cm & $>$ 5.0 cm\\
    $\vert(r_{\Omega}-r_{PV})\times p_{\Omega}\vert$/$\vert r_{\Omega}-r_{PV}\vert\vert{p_{\Omega}}\vert$ & $<$ 0.05 & $<$ 0.08 & $<$ 0.15 \\
    $DL(\Omega) < DL(\Lambda)$ & Yes & Yes & Yes \\
    proton DCA & $>$ 0.8 cm & $>$ 0.8 cm & $>$ 0.6 cm \\
    pion DCA & $>$ 2.0 cm & $>$ 2.0 cm & $>$ 1.8 cm \\
    bachelor DCA & $>$ 1.2 cm & $>$ 1.2 cm & $>$ 1.0 cm \\
    proton to pion DCA & $<$ 0.8 cm & $<$ 0.8 cm & $<$ 1.0 cm \\
    $\Lambda$ DCA to bachelor & $<$ 0.8cm & $<$ 0.8cm & $ <$ 1.0cm\\
    $\vert M_{\Lambda}-1.1156\vert$ GeV/\it{c}$^{2}$ & $<$ 0.007 GeV/\it{c}$^{2}$ &$<$ 0.007 GeV/\it{c}$^{2}$ &$<$ 0.007 GeV/\it{c}$^{2}$\\
     $\vert M_{\Omega}-1.672\vert$ GeV/\it{c}$^{2}$ & $<$ 0.007 GeV/\it${c}^{2}$ &$<$ 0.007 GeV/\it{c}$^{2}$ &$<$ 0.007 GeV/$\it{c}^{2}$\\

    \hline
  \end{tabular}
  
  \caption{\label{tab:cuts} Selection criteria for $\Omega$ and $\bar{\Omega}$ reconstruction.}

\end{table*}

\subsection{PROTON IDENTIFICATION}

\begin{figure}[h]
\begin{center}

\epsfxsize = 3.1in
\epsfysize = 2.9in
\epsffile{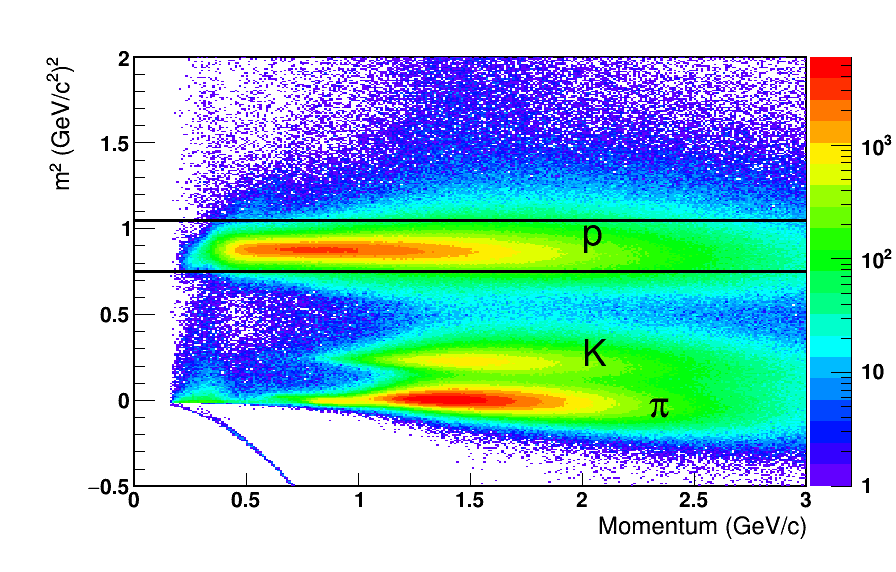} 
\end{center}
\caption{(color online)  Proton identification using the time of flight and momentum from the TOF and the TPC detectors, respectively. The solid lines show lower and upper cuts to select protons.}
\label{fig:ProtonID}
\end{figure}

The TOF and TPC detectors were used for proton (antiproton) identification in the pseudorapidity range $|\eta|< 1$. The proton tracks were selected if their DCA  was less than 0.5~cm to the primary vertex, greater than 20 points were measured out of a maximum of 45, and the number of points used in track reconstruction divided by the number of possible points was greater than 0.52 in order to prevent split tracks. The time of flight of the particles reaching the TOF detector along with the tracking information from the TPC detector was used to calculate the square of the particle mass ($m^{2}$) to identify protons. Figure~\ref{fig:ProtonID} shows  $m^{2}$ from the TOF detector versus momentum from the TPC.  All candidates with $m^{2}$ between 0.75 and 1.10 (GeV$/{\it{c}}^2)^{2}$ were used in the analysis.

\subsection{TWO-PARTICLE CORRELATION FUNCTION}

The two-particle correlation function is defined as: 

\label{myeq}
\begin{equation}
C_{\rm{measured}}({\textit{k}}^{\ast}) = \frac{A({\textit{k}}^{\ast})}{B({\textit{k}}^{\ast})},
\end{equation}

\noindent
where $A({\textit{k}}^{\ast})$ is the distribution of the invariant relative momentum, where ${\textit{k}}^{\ast} = |\vec{{\textit{k}}^{\ast}}|$ is relative momentum of one of the particles in the pair rest frame, for a proton and $\Omega$ pair or anti-proton and $\bar{\Omega}$ pair from the same event. $B({\textit{k}}^{\ast})$ is the reference distribution generated by mixing particles from different events with the same centrality and with approximately the same vertex position along the $z$-direction. The same single- and pair-particle cuts were applied for real and mixed events.  The data analysis was done in nine centrality bins: 0-5\%, 5-10\%, 10-20\%, 20-30\%, 30-40\%, 40-50\%, 50-60\%, 60-70\% and 70-80\% for both same events and mixed events. The final results were combined and presented in two centrality bins: 0-40\% and 40-80\%. The efficiency and acceptance effects canceled out in the ratio $A({\textit{k}}^{\ast})/B({\textit{k}}^{\ast})$. Corrections to the raw correlation functions were applied according to the expression:

\label{myeq2}
\begin{equation}
C'({\textit{k}}^{\ast}) = \frac{C_{\rm{measured}}({\textit{k}}^{\ast})-1}{P({\textit{k}}^{\ast})}+1, 
\end{equation}
\noindent 
where the pair purity, $P({\textit{k}}^{\ast})$, was calculated as a product of $S/(S+B)$ for the $\Omega$ ($\bar{\Omega}$) and purity of the proton (antiproton). The selected sample of proton candidates also included secondary protons from $\Lambda$, $\Sigma$ and $\Xi$ decays. The estimated fraction of primary protons (antiprotons) from thermal model~\cite{therminator} studies is 52\% (48\%)~\cite{STAR_Plm}. The purity of the proton sample is obtained as a product of identification probability and fraction of primary protons. The pair purity is 0.2 (0.36) for 0-40\% (40-80)\% centrality and  is constant over the analyzed range of invariant relative momentum. 

\begin{figure*}
\begin{center}

\epsfxsize = 6.0in
\epsfysize = 2.9in
\epsffile{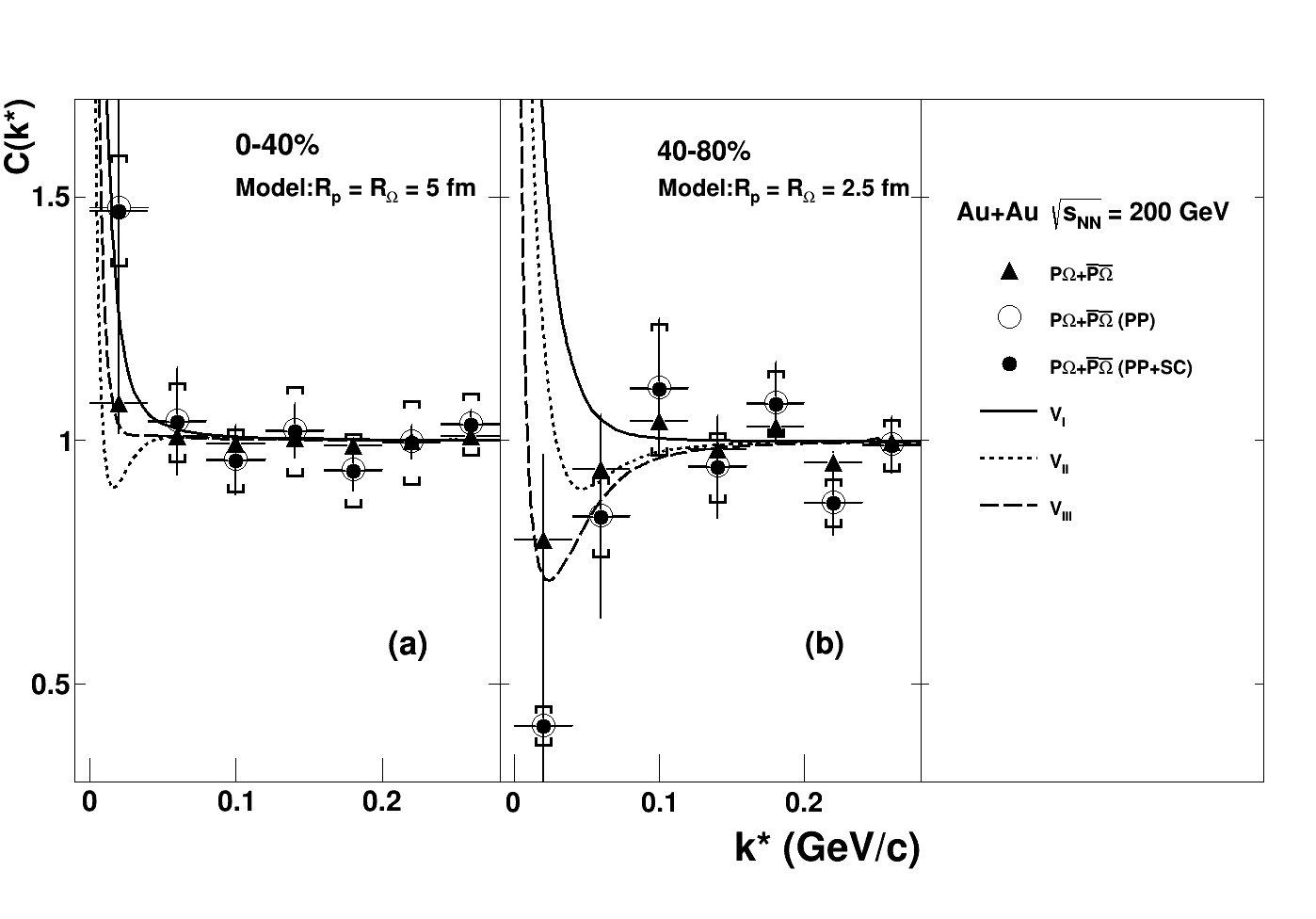} 
\end{center}
\caption{Measured correlation function (C(${\textit{k}}^{\ast}$)) for proton-$\Omega$ and antiproton-$\bar{\Omega}$ (P$\Omega+\bar{\normalfont{P}}\bar{\Omega}$) for (0-40)$\%$ (a) and (40-80)$\%$ (b) Au+Au collisions at \sqrtsNN\,\,=200 GeV. The triangles represent raw correlations, open circles represent pair-purity corrected (PP) correlations, and solid circles represent pair-purity and smearing corrected (PP+SC) correlations. The error bars correspond to statistical errors and caps correspond to the systematic errors. The predictions from~\cite{KM_AO} for proton-$\Omega$ interaction potentials $V_{I}$, $V_{I\hspace{-0.1em}I}$ and $V_{I\hspace{-0.1em}I\hspace{-0.1em}I}$ for source sizes $R_{p}$ = $R_{\Omega}$ = 5 fm  and  $R_{p}$ = $R_{\Omega}$ = 2.5 fm are shown in (a) and (b) respectively.}
\label{fig:CFcentrality}
\end{figure*}

The effect of momentum resolution on the correlation functions has also been investigated using simulated tracks from $\Omega$ decay and tracks for protons, with known momenta, embedded into real events. Correlation functions have been corrected for momentum resolution using the expression:
\begin{equation}
C({\textit{k}}^{\ast}) = \frac{C'({\textit{k}}^{\ast}) C_{\rm in}({\textit{k}}^{\ast})}{C_{\rm res}({\textit{k}}^{\ast})}, 
\end{equation}
\noindent 
where $C({\textit{k}}^{\ast})$ represents the corrected correlation function, and $C_{\rm in}({\textit{k}}^{\ast}) / C_{\rm res}({\textit{k}}^{\ast})$ is the correction factor. $C_{\rm in}({\textit{k}}^{\ast})$ was calculated without taking into account the effect of momentum resolution and $C_{\rm res}({\textit{k}}^{\ast})$ included the effect of momentum resolution applied to each $\Omega$ and proton candidates.  More details related to these corrections can be found in Ref.~\cite{pionHBT}. The impact of momentum resolution on correlation functions is negligible compared with statistical errors.

To study the shape of correlation function for the background, the candidates from the side-bands of invariant mass of $\Omega$ are chosen in the range M$<$1.665 GeV/${\it{c}}^{2}$ and M$>$1.679 GeV/${\it{c}}^{2}$. These selected candidates are then combined with the proton tracks from the same event to construct the relative momentum for the same event. The relative momentum for the mixed event is generated by combining the selected candidates from the side-band of invariant mass of $\Omega$ with protons from different events with approximately the same vertex position along the z-direction.

\section{RESULTS AND DISCUSSION}

After applying the selection criteria for proton and $\Omega$ identification, as mentioned in the data analysis section, a total of 38065$\pm$195 (8816$\pm$94) and 3037$\pm$55 (679$\pm$26) pairs of proton-$\Omega$ and  antiproton-$\bar{\Omega}$ for ${\it{k}}^{\ast}<$0.2 (0.1) GeV/{\it{c}} are observed for (0-40)\% and (40-80)\% Au+Au collisions, respectively. The measured proton-$\Omega$ and antiproton-$\bar{\Omega}$ correlation functions, P$\Omega+\bar{\normalfont{P}}\bar{\Omega}$, the correlation functions after corrections for pair purity, $P\Omega+\bar{\normalfont{P}}\bar{\Omega}$ (PP), and the correlation function after corrections for pair purity and momentum smearing, $P\Omega+\bar{\normalfont{P}}\bar{\Omega}$ (PP+SC), for 0-40\% and 40-80\% Au+Au collisions at $\sqrt{s_{\normalfont{NN}}} = 200$ GeV are shown in Figures~\ref{fig:CFcentrality} (a) and~\ref{fig:CFcentrality} (b). The systematic errors for the measured proton-$\Omega$ correlation function were estimated by varying the following requirements for the selection of $\Omega$ candidates: the decay length, DCA of $\Omega$ to the primary vertex, pointing angle cuts and mass range, which affect the purity of the $\Omega$ sample. The DCA and $m^{2}$ requirements were varied to estimate systematic from the proton purity. In addition, systematic errors from normalization and feed-down contributions were also estimated. The systematic errors from different sources were then added in quadrature. The combined systematic errors are shown in Figure~\ref{fig:CFcentrality} as caps for each bin of the correlation function.

Predictions for the proton-$\Omega$ correlation function from~\cite{KM_AO} for proton-$\Omega$ interaction potentials $V_{I}$, $V_{I\hspace{-0.1em}I}$ and $V_{I\hspace{-0.1em}I\hspace{-0.1em}I}$ for a static source with sizes $R_{p}$ = $R_{\Omega}$ = 5.0 fm and $R_{p}$ = $R_{\Omega}$ = 2.5 fm are also shown in the Figure~\ref{fig:CFcentrality}(a) and Figure~\ref{fig:CFcentrality}(b). The selected source sizes are not fit to the experimental data.  The choice of the potentials in Ref.~\cite{KM_AO} is based on an attractive N$\Omega$ interaction in the $^{5}S_{2}$ channel from the lattice QCD simulations with heavy {\it u-, d-, s-} quarks from Ref.~\cite{Nomega_BE2}.  The potential $V_{I\hspace{-0.1em}I}$ is obtained by fitting the lattice QCD data with a function $V(r)=b_{1}e^{-b_{2}r^{2}}+b_{3}(1-e^{-b_{4}r^{2}})(e^{-b_{5}r}/r)^{2}$, where $b_{1}$ and $b_{3}$ are negative and $b_{2}$, $b_{4}$ and $b_{5}$ are positive, which represents a case with shallow N$\Omega$ bound state. Two more potentials $V_{I}$ and $V_{I\hspace{-0.1em}I\hspace{-0.1em}I}$  represent cases without a N$\Omega$ bound state and  a deep N$\Omega$ bound state, respectively. Binding energy ($\bf{E_{\normalfont{b}}}$), scattering length ($\bf{a_{0}}$) and effective range ($\bf{r_{\normalfont{eff}}}$) for the N$\Omega$ interaction potentials $V_{I}$, $V_{I\hspace{-0.1em}I}$ and $V_{I\hspace{-0.1em}I\hspace{-0.1em}I}$ are listed in Table~\ref{tab:scatt_eff}~\cite{KM_AO}. The measured correlation functions for $P\Omega+\bar{\normalfont{P}}\bar{\Omega}$ are in agreement with the predicted trend for the $P\Omega$ correlation functions with interaction potentials  $V_{I}$, $V_{I\hspace{-0.1em}I}$ and $V_{I\hspace{-0.1em}I\hspace{-0.1em}I}$ for the 0-40\% Au+Au collisions as shown in Figure~\ref{fig:CFcentrality}(a). However, due to limited statistics at lower ${\it{k}}^{\ast}$, strong enhancement due to Coulomb interaction is not visible in the 40-80\% Au+Au collisions in Figure~\ref{fig:CFcentrality}(b).

\begin{table}[hbt]
  \centering
  \begin{tabular}{ c | c c c }
    \hline
    Spin-2 p$\Omega$ potentials  & $V_{I}$ & $V_{I\hspace{-0.1em}I}$ & $V_{I\hspace{-0.1em}I\hspace{-0.1em}I}$ \\
    \hline
    $\bf{E_{\normalfont{b}}}$ (MeV) & - & 6.3  & 26.9 \\
    $\bf{a_{0}}$ (fm) & -1.12 & 5.79  & 1.29 \\
    $\bf{r_{\normalfont{eff}}}$ (fm) & 1.16 & 0.96  & 0.65 \\    
    \hline
  \end{tabular}
  
  \caption{\label{tab:scatt_eff} Binding energy ($\bf{E_{\normalfont{b}}}$), scattering length ($\bf{a_{0}}$) and effective range ($\bf{r_{\normalfont{eff}}}$) for the Spin-2 proton-$\Omega$ potentials~\cite{KM_AO}.}

\end{table}


The measured proton-$\Omega$ and antiproton-$\bar{\Omega}$ correlation function includes three effects coming from the elastic scattering in the $^{5}S_{2}$ channel, the strong absorption in the $^{3}S_{1}$ channel and the long-range Coulomb interactions.  The Coulomb interaction between the positively charged proton and negatively charged $\Omega$ introduces a strong enhancement in the correlation function at small ${\textit{k}}^{\ast}$, as seen in Figure~\ref{fig:CFcentrality}. One can remove the Coulomb enhancement using a Gamow correction, however, this simple correction is not good enough to extract the characteristic feature of correlation function from strong interaction. A full correction with source-size dependence is needed to isolate the effect of strong interaction from Coulomb enhancement. Therefore the ratio of correlation function between small and large collision systems, is proposed in \cite{KM_AO} as a model-independent way to access the strong interaction with less contamination from the Coulomb interaction.

\begin{figure*}
\begin{center}
\epsfxsize = 6.0in
\epsfysize = 4.0in
\epsffile{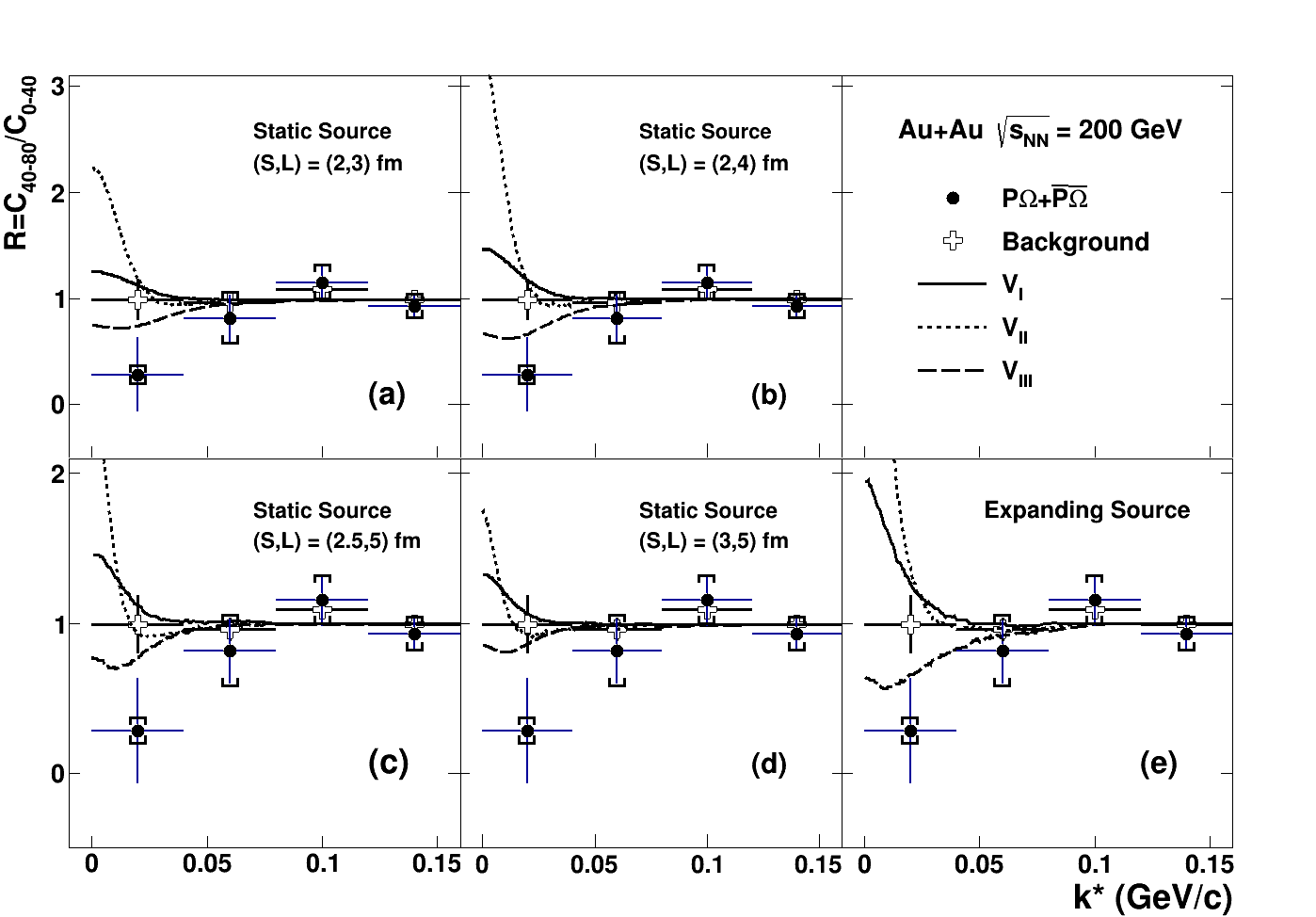} 
\end{center}
\caption{The solid circle represents the ratio (R) of small system (40-80$\%$ collisions) to large system (0-40$\%$ collisions) for 
proton-$\Omega$ and antiproton-$\bar{\Omega}$ ($P\Omega+\bar{\normalfont{P}}\bar{\Omega}$). The error bars correspond to the statistical errors and caps correspond to systematic errors. The open crosses represent the ratio for background candidates from the side-band of $\Omega$ invariant mass.  Predictions for the ratio of small system to large system~\cite{KM_AO, KM_priv} for proton-$\Omega$ interaction potentials $V_{I}$ , $V_{I\hspace{-0.1em}I}$  and $V_{I\hspace{-0.1em}I\hspace{-0.1em}I}$ for static source with different source sizes (S,L) = (2,3), (2,4), (2.5, 5) and (3,5) fm, where S and L corresponding to small and large systems, are shown in (a), (b), (c) and (d) respectively. In addition, the prediction for the expanding source is shown in (e). }
\label{fig:latticedata}
\end{figure*}

The ratio of combined proton-$\Omega$ and antiproton-$\bar{\Omega}$ correlation function from peripheral (40-80$\%$) to central (0-40$\%$) collisions, defined as R = C$_{40-80}$/C$_{0-40}$, for proton-$\Omega$ and antiproton-$\bar{\Omega}$ ($P\Omega+\bar{\normalfont{P}}\bar{\Omega}$) is shown in Figure~\ref{fig:latticedata}. The systematic uncertainties are propagated from the measured correlation functions for 0-40$\%$ and 40-80$\%$ centrality and are shown as caps. For the background study, the candidates from the side-bands of $\Omega$ invariant mass were combined with protons to construct the correlation function. The same ratio, R, for the background is unity and is shown as open crosses in Figure~\ref{fig:latticedata}. Previous measurements of source size for $\pi$-$\pi$, $K^{0}_{S}$-$K^{0}_{S}$, proton-proton and proton-$\Lambda$ correlations show that the source size decreases as the transverse mass increases~\cite{pbar_STAR, pionHBT,STAR_Plm,KKHBT, Hannathesis}. Using this  transverse mass dependence~\cite{Hannathesis}, the expected source size for proton-$\Omega$ is 2-3 fm for peripheral collisions and 3-5 fm for central collisions. The predictions for the ratio of small system to large system from Refs.~\cite{KM_AO, KM_priv} for proton-$\Omega$ interaction potentials $V_{I}$, $V_{I\hspace{-0.1em}I}$ and $V_{I\hspace{-0.1em}I\hspace{-0.1em}I}$ for static source with different source sizes (S,L) = (2,3), (2,4), (2.5, 5) and (3,5) fm, where S and L correspond to small and large collision sytem, are shown in Figure~\ref{fig:latticedata}(a-d).  A small variation in the source size does not change the characteristic of the ratio for the choice of three potentials.

Predictions for the ratio of small to large system  with the effects of collective expansion are also shown in the Figure~\ref{fig:latticedata}(e)~\cite{KM_AO}. The transverse source sizes are taken as $R^{\normalfont{tr}}_{\normalfont{p}}$ = $R^{\normalfont{tr}}_{\Omega}$ = 2.5 fm  for small system and $R^{\normalfont{tr}}_{\normalfont{p}}$ = $R^{\normalfont{tr}}_{\Omega}$ = 5 fm for large system. The temperature at the thermal freeze-out is $T_{\normalfont{p},\Omega}$ = 164 MeV for peripheral collisions and $T_{\normalfont{p},\Omega}$ = 120 MeV for the central collisions~\cite{Abelev, Shen} and the proper-time at the thermal freeze-out is $\tau_{\normalfont{p}} (\tau_{\Omega})$ = 3(2) fm/\textit{c} for the peripheral collisions and  $\tau_{\normalfont{p}} (\tau_{\Omega})$ = 20(10) fm/\textit{c} for the central collisions~\cite{Zhu}. 

The predictions with expanding source for the proton-$\Omega$ interaction potentials $V_{I}$ and $V_{I\hspace{-0.1em}I}$ are 3$\sigma$ larger than the data at ${\textit{k}}^{\ast} =20$ MeV/\textit{c}. The prediction for the proton-$\Omega$ interaction potential $V_{I\hspace{-0.1em}I\hspace{-0.1em}I}$ with expanding source and static source are within 1$\sigma$ of the data at ${\textit{k}}^{\ast} =20$ MeV/\textit{c}. As shown in Figure~\ref{fig:latticedata}, the measured ratios at  ${\textit{k}}^{\ast}$ =20 and 60 MeV/\textit{c} are R$= 0.28\pm0.35_{\normalfont{stat}}\pm0.03_{\normalfont{sys}}$ (background = $0.96\pm0.13_{\normalfont{stat}})$ and R$=0.81\pm0.22_{\normalfont{stat}}\pm0.08_{sys}$ (background = $0.97\pm0.05_{\normalfont{stat}})$, respectively. Comparing these values with the model calculations shown in Figure 5(b) of the Ref.~\cite{KM_AO}, where a bound state with  $\bf{E_{\normalfont{b}}}\sim$27 MeV for the proton-$\Omega$ system is assumed in calculation, we conclude that our data favor a positive scattering length for the proton-$\Omega$ interactions. The positive scattering length and the measured ratio of proton-$\Omega$ correlation function from peripheral to central collisions less than unity for ${\textit{k}}^{\ast} < 40$ MeV/\textit{c} favors the proton-$\Omega$ interaction potential $V_{I\hspace{-0.1em}I\hspace{-0.1em}I}$ with $\bf{E_{\normalfont{b}}}\sim$27 MeV for proton and $\Omega$.

\section{CONCLUSIONS}
The first measurement of the proton-$\Omega$ correlation function in heavy-ion collisions for Au+Au collisions at \sqrtsNN\,\,=200 GeV is presented in this Letter. The measured ratio of proton-$\Omega$ correlation function from peripheral to central collisions is compared with the predictions based on proton-$\Omega$ interaction extracted from (2+1)-flavor lattice QCD simulations. At present, due to limited statistics, it is not possible to extract the interaction parameters.  However the measured ratio of proton-$\Omega$ correlation function from peripheral to central collisions less than unity for ${\textit{k}}^{\ast}<40$ MeV/\textit{c} within 1$\sigma$ indicates that the scattering length is positive for the proton-$\Omega$ interaction and favors the proton-$\Omega$ bound state hypothesis.  

\section{ACKNOWLEDGMENTS}

We thank Dr. Kenji Morita, Dr. Akira Ohnishi, Dr. Faisal Etminan and Dr. Tetsuo Hatsuda for providing the calculation and enlightening discussions. We thank the RHIC Operations Group and RCF at BNL, the NERSC Center at LBNL, and the Open Science Grid consortium for providing resources and support. This work was supported in part by the Office of Nuclear Physics within the U.S. DOE Office of Science, the U.S. National Science Foundation, the Ministry of Education and Science of the Russian Federation, National Natural Science Foundation of China, Chinese Academy of Science, the Ministry of Science and Technology of China  (973 Program No. 2014CB845400, 2015CB856900) and the Chinese Ministry of Education, the National Research Foundation of Korea, Czech Science Foundation and Ministry of Education, Youth and Sports of the Czech Republic, Department of Atomic Energy and Department of Science and Technology of the Government of India, the National Science Centre of Poland, the Ministry of Science,  Education and Sports of the Republic of Croatia, RosAtom of Russia and German Bundesministerium fur Bildung, Wissenschaft, Forschung and Technologie (BMBF) and the Helmholtz Association.

\renewcommand{\bibfont}{\small}

\end{document}